\documentclass[twocolumn]{emulateapj}
\usepackage{multirow}
\usepackage{graphicx}
\usepackage{footnote}
\usepackage{natbib}
\usepackage[fleqn]{amsmath}
\usepackage{amsmath}
\usepackage{color}
\usepackage{subfigure}
\usepackage{comment}

\usepackage{ulem}

\shorttitle{VLBI observations of PTF11qcj}
\shortauthors{Palliyaguru et al.}

\begin{document}

\title{VLBI observations of supernova PTF11qcj: \\Direct constraints on the size of the radio ejecta}

\author{N. T. Palliyaguru \altaffilmark{1,*}}
\author{A. Corsi \altaffilmark{1}}
\author{M. P\'erez-Torres \altaffilmark{2}}
\author{E. Varenius \altaffilmark{3,4}}
\author{H. Van Eerten\altaffilmark{5}}
\altaffiltext{1}{Department  of  Physics  and  Astronomy,  Texas  Tech  University,  Lubbock,  TX  79409-1051  (USA)}
\altaffiltext{$\star$}{Email: \email{Nipuni.Palliyaguru@ttu.edu}}
\altaffiltext{2}{Instituto de Astrofísica de Andalucía -Consejo Superior de Investigaciones Científicas (CSIC), PO Box 3004, 18008, Granada, Spain.}
\altaffiltext{3}{Department of Earth and Space Sciences, Chalmers University of Technology, Onsala Space Observatory, 439 92 Onsala, Sweden.}
\altaffiltext{4}{Jodrell Bank Centre for Astrophysics, The University of Manchester, Oxford Rd, Manchester M13 9PL, UK.}
\altaffiltext{5}{Department of Physics, University of Bath, Claverton Down, Bath BA2 7AY, UK}

\begin{abstract}
 We present High Sensitivity Array (HSA) and enhanced Multi Element Remotely Linked Interferometer Network (eMERLIN) observations of the radio-loud broad-lined type Ic supernova PTF11qcj obtained $\sim7.5$ years after the explosion. Previous observations of this supernova at 5.5\,yrs since explosion showed a double-peaked radio light curve accompanied by a detection in the X--rays, but no evidence for broad H$\alpha$ spectral features.
The Very Long Baseline Interferometry (VLBI) observations presented here show that the PTF11qcj GHz radio ejecta remains marginally resolved at the sub-milliarcsecond level at $\approx 7.5$\,yrs after the explosion, pointing toward a non-relativistic expansion. Our VLBI observations  thus favor a scenario in which the second peak of the  PTF11qcj radio light curve is related to strong interaction of the supernova ejecta with a circumstellar medium of variable density, rather than to the emergence of an off--axis jet. Continued VLBI monitoring of  PTF11qcj in the radio may strengthen further this conclusion.
\end{abstract}

\keywords{}

\section{Introduction} 
\label{sec:intro}
Supernovae (SNe) of type Ib/c are believed to mark the deaths of massive stars that are stripped of their hydrogen (type Ib), and possibly helium (type Ic), envelope before explosion \citep{Filippenko1997}. A sub-class of Ib/c SNe dubbed broad-line (BL) Ic, estimated to constitute only $\approx 5\%$
of the Ib/c population \citep{Woosley2006,Gal-Yam2016}, is of particular interest due to its relation to long-duration gamma-ray bursts (GRBs), the most relativistic stellar explosions we know of in the universe \citep{Piran2004,Meszaros2006}.
While all GRB-associated SNe are of type BL-Ic \citep[][ but see \citet{Cano2014} for the peculiar case of SN\,2013ez]{Woosley2006, hb12}, not all BL-Ic events make a GRB \citep[e.g.,][]{Berger2003,Soderberg2006,Corsi2016}. Thus, the question of what physical ingredients enable some stripped-envelope massive stars to launch a relativistic jet remains open \citep[e.g.,][]{Modjaz2016}.

As first demonstrated by the well-known case of SN\,1998bw/GRB\,980425 \citep{Galama1998,Kulkarni1998}, radio observations are particularly well suited to identify those BL-Ic SNe that may harbor GRBs (hereafter referred to as engine-drive SNe), since radio synchrotron emission traces the fastest moving ejecta \citep[e.g.,][]{Berger2003}. At the same time, because non-thermal radio  photons are produced in the interaction of the SN shock with the circumstellar material (CSM), bright radio emission can also be the smoking gun for (non-relativistic) ejecta interacting with a high-density CSM \citep[e.g.,][]{Chev98, Chev04, Chev06}. Although strong CSM interaction is not commonly observed in BL-Ic SNe, a few cases exists such as SN\,2007bg \citep{Salas2013}, PTF11qcj \citep{Corsi2014,Palliyaguru2019}, SN2018gep \citep{Ho2019a} and, possibly, AT2018cow \citep{Rivera2018,Smartt2018,Ho2019,Margutti2019}.

Here, we focus on PTF11qcj, a radio-loud BL-Ic SN extensively monitored via our approved programs on the Karl G. Jansky Very Large Array \citep[VLA;][]{Corsi2014,Palliyaguru2019}. The extraordinary radio luminosity of PTF11qcj ($\sim 10^{29} \rm {erg\,s^{-1}\,Hz^{-1}}$) is reminiscent of the GRB--associated SN\,1998bw \citep{Kulkarni1998}. As discussed in \citet{Corsi2014} and \citet{Palliyaguru2019}, our radio monitoring over the first $\approx 5.5$ years since explosion has revealed an unusual double--peaked radio light curve. The radio emission observed during the first light curve peak ($t\lesssim 200$\,d) can be modeled within the standard synchrotron self-absorption (SSA) model for a spherical SN shock expanding in the CSM. This model yields an estimated speed of $\approx 0.3-0.5\,c$ for the fastest SN ejecta \citep{Corsi2014}, placing PTF11qcj in an intermediate class between ``ordinary'' BL-Ic SNe and engine-driven ones like SN\,1998bw or SN\,2009bb \citep{Soderberg2010}. The simple, spherically symmetric model of SN shock interaction with a smooth CSM (simple power-law density profile), however, cannot explain the second radio peak. As discussed in \citet{Palliyaguru2019}, two more complex scenarios can be invoked to interpret this peculiar behavior of PTF11qcj:
 (i) A spherical SN shock going through a medium with extreme CSM density variations, perhaps related to eruptive progenitor mass loss; (ii) A radio-emitting SN shock (first peak) followed by radio emission from an emerging off-axis GRB jet, initially pointed away from our line of sight (second peak).

In \citet{Palliyaguru2019} we have shown that, while modeling of our VLA dataset within scenario (i) can indeed explain the second radio peak, the presence of an off--axis jet (scenario (ii)) cannot be ruled out just based on light curve measurements. However, scenarios (i) and (ii) make rather different predictions for the angular size of the PTF11qcj ejecta at very late times. 
Motivated by these considerations, here we present Very Long Baseline Interferometry (VLBI) observations of PTF11qcj aimed at setting direct constraints on the size (angular diameter) of its radio ejecta. These observations ultimately provide a direct test for the presence of relativistic expansion, as expected in the case of an off-axis GRB jet. 

Our paper is organized as follows. In Section~\ref{sec:obs}, we present the HSA and eMERLIN observations of PTF11qcj. In Section~\ref{sec:models}, we discuss these observations within the light curve and radio ejecta size predictions of the two scenarios mentioned above. Finally, in Section~\ref{sec:conc} we summarize our results and conclude. 

\section{Observations and data reduction}
\label{sec:obs}

\begin{table*}
\centering
\caption{HSA results for PTF11qcj (project codes BP229A and BP229B), see Section {\ref{sec:vlba_obs}} for discussion.}
\label{tab:hsa}
\begin{tabular}{ l | c c} 
\hline\hline
 & 1.66~GHz  & 15.37~GHz  \\ 
 \hline
 Observing date (UT) & 2018-12-08 & 2019-04-28 \\ 
 Project code & BP229A & BP229B \\  
 Observing time including calibrators (h)& 8 & 8 \\ 
 Image off-source RMS noise ($\mu$Jy/beam)& 125 & 24 \\
 CLEAN restoring beam (mas)$^2$ & $4.06\times2.66$ & $0.54\times0.24$ \\ 
 CLEAN beam position angle (deg) & -29 & -21 \\ 
 \hline
 Peak flux density (mJy) & $4.32\pm0.92$ & $0.494\pm0.054$\\ 
 Integrated flux density (mJy) & $5.8\pm1.4$ & $0.681\pm0.085$ \\ 
 Right Ascension [J2000]& 13$^{\rm h}$13$^{\rm m}$41.47512$^{\rm s}$ & 13$^{\rm h}$13$^{\rm m}$41.47490$^{\rm s}$ \\ 
 Declination [J2000]& +47$^\circ$17$'$56.8017$''$ & +47$^\circ$17$'$56.7988$''$ \\
 \hline
 Deconvolved fitted major axis & - & $300\pm71\,\mu$arcsec \\
Deconvolved fitted minor axis & - & $76\pm76\,\mu$arcsec \\ 
Deconvolved fitted Pos. ang. & - & $125\pm19^\circ$ \\
\hline
\end{tabular}
\end{table*}

\begin{table*}
\centering
\caption{\lowercase{e}MERLIN results for PTF11qcj (project code DD8011), see Section {\ref{sec:emerlin_obs}} for discussion. }
\label{tab:emerlin}
\begin{tabular}{ l | c  c } 
\hline\hline
 & 1.5~GHz & 5.07~GHz \\ 
 \hline
 Observing date (UT) & 2019-08-29 & 2019-08-01 \\ 
 Project code & DD8011 & DD8011 \\  
  Observing time including calibrators (h)& 6 & 5 \\ 
 Image off-source RMS noise ($\mu$Jy/beam) & 42 & 37 \\ 
 CLEAN beam (mas$^2$) &$59\times19$& $622\times78$ \\
 CLEAN beam position angle (deg)& 40 & 42 \\ 
 \hline
 Peak flux density (mJy) & $5.82\pm0.87$ & $4.15\pm0.62$\\ 
 Integrated flux density (mJy) & $5.78\pm0.87$ & $4.45\pm0.67$ \\ 
 Right Ascension [J2000]& 13$^{\rm h}$13$^{\rm m}$41.4745$^{\rm s}$ & 13$^{\rm h}$13$^{\rm m}$41.4746$^{\rm s}$ \\ 
 Declination [J2000]& +47$^\circ$17$'$56.795$''$ & +47$^\circ$17$'$56.800$''$ \\
 \hline
\end{tabular}
\end{table*}

\subsection{HSA observations} 
\label{sec:vlba_obs}
We observed the field of PTF11qcj at 1.66\,GHz (project code BP229A, PI: Palliyaguru) and 15.37\,GHz (BP229B, PI: Palliyaguru) with the High Sensitivity Array  (HSA) on 2018 December 08.37 UTC and 2019 April 28.97 UTC. The HSA included, in both bands, the Very Long Baseline Array (VLBA, USA) and the Effelsberg 100\,m antenna (Germany). At 1.66\,GHz, we also used the VLA 
for improved sensitivity. Both observations covered 128\,MHz continuum bandwidth and were correlated at the National Radio Astronomy Observatory (NRAO) Array Operations Center (AOC) in Socorro (New Mexico, USA) with averaging times of 2\,s and 1\,s at 1.66\,GHz and 15.37\,GHz, respectively. In both observations, we correlated the target data at R.A. $13^{\rm h}13^{\rm m}41.5100^{\rm s}$, Dec. $+47^\circ17'57.600''$ \citep[J2000;][]{Corsi2014}.
In setting up our observations, we used standard phase referencing, where scans on target are interleaved with scans on a nearby compact complex gain calibrator with known position. At 1.66\,GHz and 15.37\,GHz, we used J1310+4653 and J1358+4737, respectively, as our complex gain calibrators. 
The VLBA observations were correlated using the NRAO's implementation of the DiFX software correlator \citep{Deller11}.

\begin{figure}[tbp]
\centering
\subfigure{
        \includegraphics[height = 0.28\textheight]{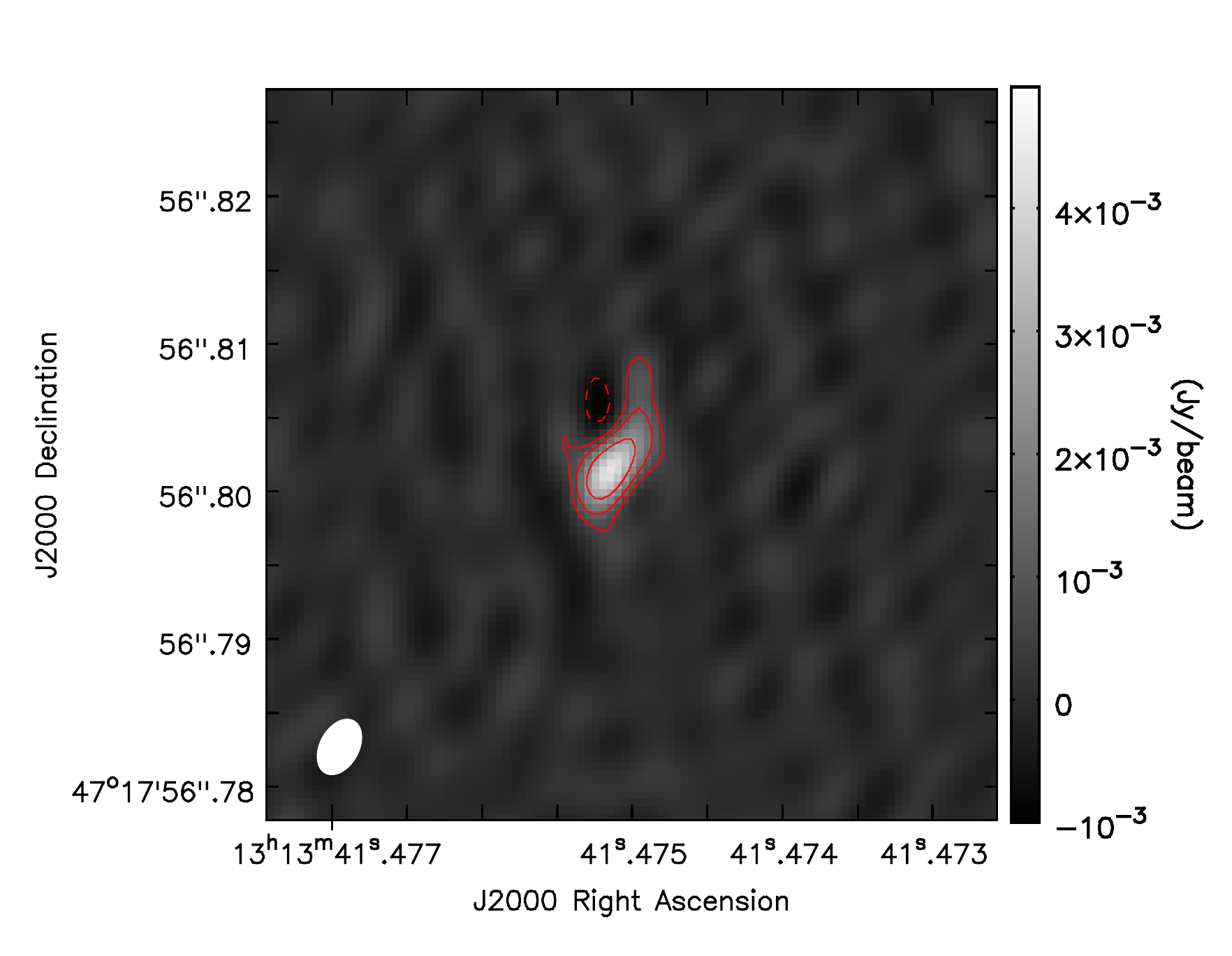}
        \label{fig:stackedeast}
}
\subfigure{
        \includegraphics[height = 0.28\textheight]{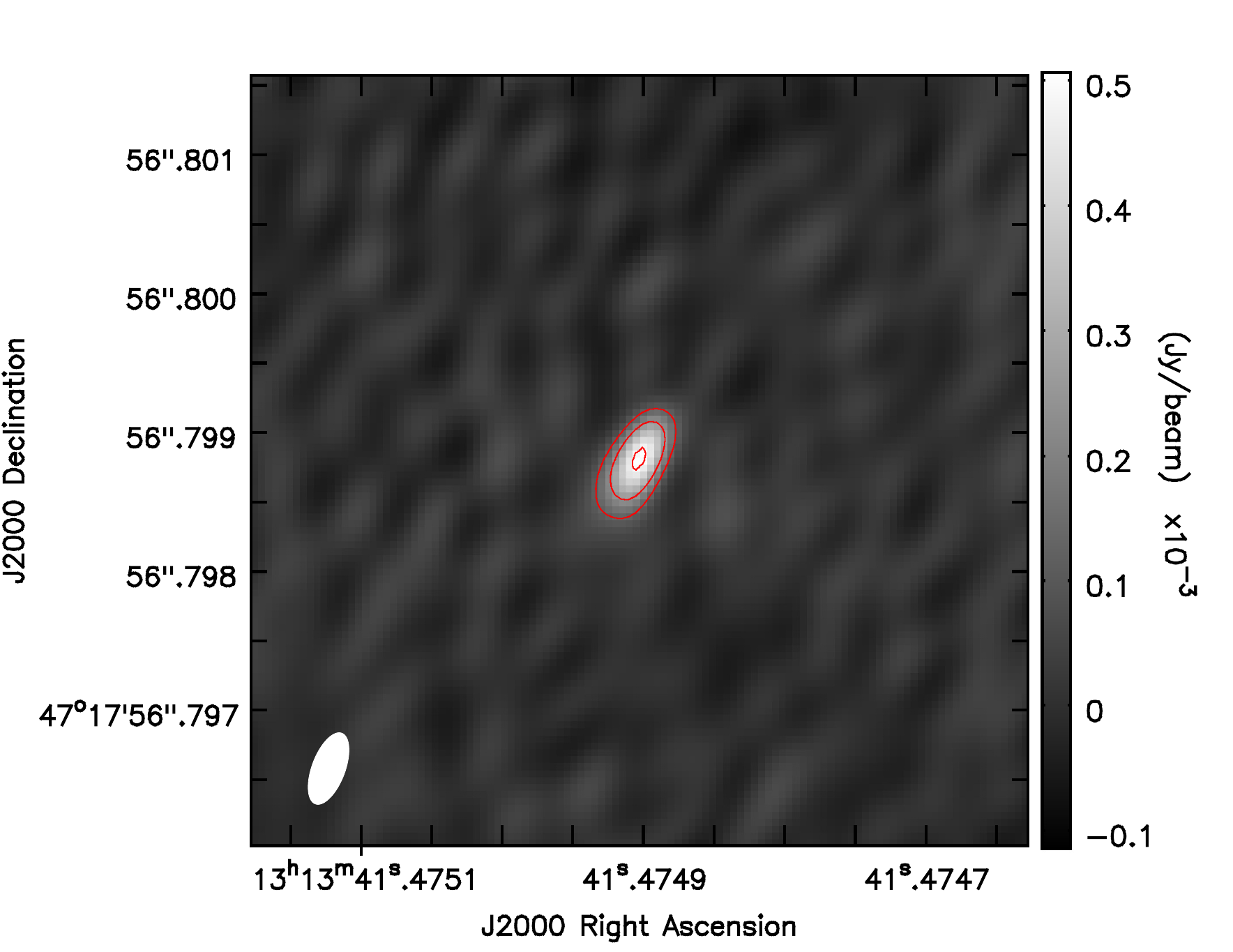}
        \label{fig:stackedeast}
}
\vspace{-0.4cm}
\caption{VLBI images of PTF11qcj at 1.66\,GHz (top) and 15.37\,GHz (bottom). Note the different scale on the axes.}
\label{fig:VLBIfigs}
\end{figure}

We performed all calibration and imaging procedures using the 31DEC19 release of the Astronomical Image Processing System  \citep[AIPS;][]{greisen2003} and ParselTongue \citep{kettenis2006}.
PTF11qcj was clearly detected at both frequencies. At 1.66\,GHz, in particular,  given the relatively large separation between PTF11qcj and the complex gain calibrator J1358+4737 ($\approx 7.6^\circ$) 
we used self-calibration to correct for the residual phase errors towards PTF11qcj and obtain a reliable flux density measurement. 
The final VLBI images are presented in Figure \ref{fig:VLBIfigs}.

We fitted Gaussian intensity profiles to the target images to obtain the flux density and position of PTF11qcj at each frequency. Our results are summarized in Table {\ref{tab:hsa}}. The reported flux density uncertainties are the quadrature sum of the instrumental calibration uncertainty plus the uncertainty of the Gaussian fit. At 15.37\,GHz, we adopt an instrumental uncertainty of 15\%. At 1.66\,GHz, we use an uncertainty of 20\% to also account for residual errors due to the large separation between  the target and the complex gain calibrator. 

To estimate the source size we fit Gaussian intensity distributions to the images. We find that the source appears marginally resolved along the major axis of our 15.37 GHz observations, while the minor axis is consistent with an unresolved source.  The deconvolved major (minor) FWHM size of the fitted Gaussian model is $300 \pm 71 \,\rm \mu arcsec$ ($76 \pm 76 \,\rm \mu arcsec$), with position angle $125 \pm 19$ deg (see also Table~\ref{tab:hsa}). This corresponds to a diameter of $(5.2\pm1.2) \times 10^{17}$ cm $((1.3\pm1.3) \times 10^{17}$ cm). Given the relatively weak radio emission from PTF11qcj,  although this results gives us an estimate of the source size, we refrain from speculation about the non-symmetrical nature of this result. PTF11qcj radio ejecta is not resolved at  1.66\,GHz.

\subsection{eMERLIN observations} 
\label{sec:emerlin_obs}
We observed the field of PTF11qcj at 5.07\,GHz and 1.51\,GHz with all six eMERLIN antennas on 2019 August 01.55 UTC and 2019 August 29.47 UTC via our DDT project DD8011 (PI: Perez-Torres). 
We used J1310+4653 as our complex gain calibrator, 1\,s integration time, and 512\,MHz continuum bandwidth in both bands. 
We used the standard eMERLIN calibrators 3C286 and OQ208 for flux density calibration and bandpass calibration, respectively. 

We calibrated and edited the correlated data using the eMERLIN CASA pipeline version 1.1.11 \citep{Moldon18}. We applied self-calibration in both bands to correct for significant residual phase errors and minor amplitude errors. We used WSClean \citep{offringa-wsclean-2014} to deconvolve the calibrated data and produce the final images. PTF11qcj is clearly detected in both bands.

We report in  Table {\ref{tab:emerlin}} the peak and total flux densities, along with the position of PTF11qcj, at each frequency, calculated by fitting Gaussian intensity profiles to the images.
The quoted flux density uncertainties correspond to the sum in quadrature of  the systematic, i.e., instrumental uncertainty (15\%) and the image off-source RMS noise.

\section{Modeling}
\label{sec:models}
\begin{figure*}
\begin{center}
\hspace{-1.3cm}
\includegraphics[width=19cm,trim={ 0.5cm 11.5cm 1cm
2.0cm},clip]{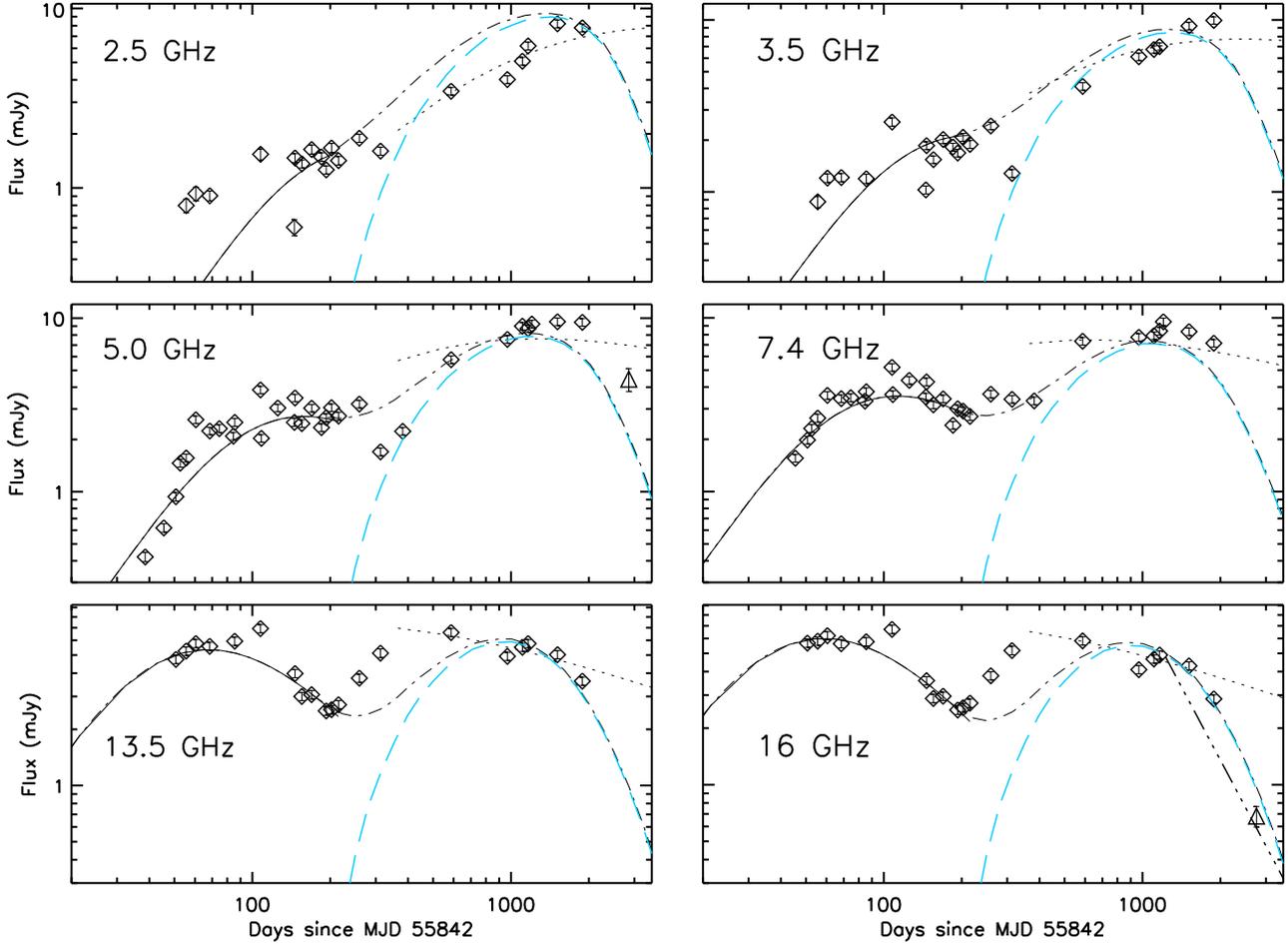}
\caption{Radio light curves of PTF11qcj obtained with the VLA, HSA and eMERLIN at six different frequencies. 
The latest HSA and eMERLIN data are shown by triangles.
The first radio peak is modeled within a standard SSA model as described in \citet{Corsi2014} (Model 0 in Table~\ref{tb:model_results_ssa}; solid curves).
The re-brightening phase is modeled in two different scenarios: (i) within the standard SSA model  (dotted curves; Model 3 in Table~\ref{tb:model_results_ssa}) and (ii) within an off-axis afterglow model  (long--dashed, blue curves; see Section \ref{sec:model_offGRB}). 
The sum of the best fit CSM model in the first peak and off-axis jet emission from the second peak is also shown (dot--dashed curve). The dash-dot-dot-dotted line in the bottom right panel shows the effects of synchrotron cooling within the SSA scenario as discussed in Section \ref{sec:model_SSA}.
The measurement of the shock size from HSA data rules out the off--axis jet scenario. 
\label{fig:lc_all_models_phase2}}
\end{center}
\end{figure*}

The complete radio light curves of PTF11qcj are shown in Figure~\ref{fig:lc_all_models_phase2}. These include the latest flux measurements (integrated fluxes from Tables~\ref{tab:hsa} and \ref{tab:emerlin}) along with data published in \citet{Corsi2016} and  \citet{Palliyaguru2019}. Hereafter, we discuss the possible interpretation of these light curves within the the standard synchrotron self-absorption (SSA) scenario for radio SNe \citep[see][and references therein]{skb05}. We also consider an alternative interpretation within an SSA (first radio light curve peak) plus off-axis GRB (second radio light curve peak) scenario. Finally, we discuss the direct size constraints obtained via our VLBI observations in the context of both these scenarios.

\begin{table*}
\begin{center}\begin{footnotesize}
\caption{Best fit parameters for the standard SSA model described in Section \ref{sec:model_SSA}.
\label{tb:model_results_ssa}}
\begin{tabular}{llllllllllll}
\hline
\hline
Parameter & Model 0 & Model 1 & Model 2 & Model 3 \\
\hline
$r_0$ (cm)& 1.1$\times10^{16}$& 1.1$\times10^{16}$ (fixed) & 1.1$\times10^{16}$ (fixed)& 1.1$\times10^{16}$ (fixed) \\
$\xi$ & 0.23 &0.19  & 0.19 &0.34\\
$\alpha_r$ & 0.79 &0.79 (fixed) & 0.79 (fixed) & 0.61\\
$t_e$& 55842 (fixed) &55842 (fixed) & 55842 (fixed) &55842 (fixed) \\
$s$ & 2.0 (fixed) & 1.4 & 1.1 & 0.0 (fixed)\\
$B_0$ (G) & 6.7 & 5.8 & 3.2& 1.35\\
$p$ & 3.0 (fixed) & 3.0 (fixed) & 3.0 (fixed) & 3.0 (fixed) \\
\hline
$\alpha_B$ & -1.0&-0.76 & -0.64 & -0.39\\
$\gamma_{m,0}$& 7.3&7.8 & 10.5 & 28.1 \\
$\alpha_{\gamma}$ &-0.4&-0.4 & -0.4 &-0.8\\
$n_{e,0}  \, (\rm cm^{-3})$ & $1.5\times10^{5}$ & $1.0\times10^{5}$ & $2.4\times10^{4}$ & $1.6\times10^{3}$\\
$\alpha_{n_e}$& -1.6 & -1.1 & 1-0.88 &0.0\\
$\dot M_0 \, (\rm M_{\odot}\, yr^{-1})$ & $1.2\times10^{-4}$ &$8.4\times10^{-5}$ & $1.9\times10^{-5}$ & $6.4\times10^{-6}$ \\
$\alpha_{\dot M}$& 0.0 &0.48 & 0.71 & 1.2\\
$\chi^2$/dof& 1793/90 &1288/39 & 602/38 & 211/38\\
\hline
\end{tabular}
\end{footnotesize}\end{center}\end{table*}
\subsection{Light curves within the SSA scenario}
\label{sec:model_SSA}
Similarly to what done in \citet{Corsi2016} and \citet{Palliyaguru2019}, we can model the radio emission of PTF11qcj in the SSA scenario \citep{skb05}. 
As described in \citet{skb05}, the temporal evolution of the shock radius, $r$, minimum Lorentz factor, $\gamma_m$, and magnetic field, $B$, are parameterized as:
\begin{eqnarray}
    r=r_0\left(\frac{t-t_e}{t_0}\right)^{\alpha_r},
    \label{eq:radius}\\
    B=B_0\left(\frac{t-t_e}{t_0}\right)^{\alpha_B},\\
    \gamma_m=\gamma_{m,0}\left(\frac{t-t_e}{t_0}\right)^{\alpha_\gamma},
\end{eqnarray}
with $\alpha_r$, $\alpha_B$ and $\alpha_\gamma$ the temporal indices of the three quantities respectively, $t_e$ the explosion time, and $t_0$ an arbitrary reference epoch since explosion, here set to $10$\,d. Within the standard assumptions, one has \citep[see Equations (9)-(10) in][]{skb05}:
\begin{eqnarray}
    \alpha_\gamma=2(\alpha_r-1),\\
    \alpha_B=\frac{(2-s)}{2}\alpha_r-1.
\end{eqnarray}
Here, $s$ describes the density profile of the shocked CSM where the density of the radiating electrons within the shocked CSM is given by \citep{Chevalier1982}
\begin{equation}
n_e=n_{e,0}\left(\frac{t-t_e}{t_0}\right)^{\alpha_{n_{e,0}}}\propto r^{-s}.
\end{equation}
The above relations follow from assuming that the energy density of shocked particles (protons and electrons) and amplified magnetic fields are a constant fraction ($\epsilon_e\approx \epsilon_B \approx 0.33$ under the hypothesis of equipartition) of the post-shock energy density $U\propto n_e v^2$, where:
\begin{equation}
    v=v_0\left(\frac{t-t_e}{t_0}\right)^{\alpha_r-1},
\end{equation}
is the shock speed.
 The density of electrons in the shocked CSM is related to the progenitor mass-loss rate via the relation \citep[see Equation (13) in][]{skb05}:
\begin{equation}
    \dot{M}=\frac{8\pi}{\eta}n_{e,0} m_p r^2_0 v_w \left(\frac{t-t_e}{t_0}\right)^{\alpha_r(2-s)}=\dot M_0 \left(\frac{t-t_e}{t_0}\right)^{\alpha_{\dot{M}}},
\end{equation}
where we have assumed a nucleon-to-proton ratio of 2, $v_w\sim 1000$\,km/s is the velocity of the stellar wind, and where $\eta$ (typically in the range $\eta \sim 2-10$) characterizes
the thickness of the radiating electron shell as $r/\eta$.

In the GHz radio band, the observing frequencies $\nu$ are typically such that $\nu_m << \nu $, where \citep[see Equation (A6) in][]{skb05}:
\begin{equation}
\nu_{m}=\gamma^2_m\left(\frac{eB}{2\pi m_e c}\right)
\end{equation}
is the characteristic synchrotron frequency of electrons with Lorentz factor $\gamma_m$. In the above Equation, $m_e$ is the electron mass, and $c$ is the speed of light. In this frequency range, assuming the synchrotron cooling frequency is higher than the observing frequency, and neglecting synchrotron cooling effects, the SSA emission from the shocked electrons reads:
\begin{equation}
    f_{\nu}(t)= {\cal F} \left(\frac{t-t_e}{t_0}\right)^{(4\alpha_r-\alpha_B)/2}(1-e^{-\tau^{\xi}_{\nu}(t)})^{1/\xi}\nu^{5/2},
\end{equation}
where ${\cal F}$ is a normalization constant that depends on the parameters $(r_0,B_0,p)$ with $p$ the power-law index of the electron energy distribution \citep[see Equations (A11)\footnote{The functions defined in (A11) are not time--dependent and therefore included in the normalization constant} and (A13) in][]{skb05}, and where the optical depth $\tau$ is given by \citep[see Equations (20) and (A14) in][respectively]{Chevalier2003,skb05}:
\begin{equation}
\tau(t)= {\cal T}\left(\frac{t-t_e}{t_0}\right)^{(p-2)\alpha_\gamma+(3+p/2)\alpha_B+\alpha_r}\,\nu^{-(p+4)/2},
\end{equation}
with ${\cal T}$ a normalization constant that depends on the parameters $(r_0,B_0,p,\gamma_{m,0},\eta)$. We note that for $\nu_m << \nu$, the value of $\gamma_{m,0}$ is left largely unconstrained by the observations,  and thus typically fixed so that $\nu_{m,0}\sim 1$\,GHz. In addition, the thickness of the shell is typically set to a value $\eta > 1$ \citep{Li1999}. Thus, for a given choice of $\eta$ and $\nu_m$, the SSA model is a function of the parameters ($r_0,\xi,\alpha_r,t_e,s,B_0,p)$.

Within the SSA scenario, and with the data collected here, we can further test the hypothesis first presented in \citet{Palliyaguru2019} that the double-peaked radio light curve of PTF11qcj is due to strong interaction with CSM of variable density. 
Our results are shown in Figure~\ref{fig:lc_all_models_phase2}, and are  reported in Table \ref{tb:model_results_ssa}.
Model 0 in Table~\ref{tb:model_results_ssa} is the best fit SSA model for the first peak ($t\lesssim 215$ days since explosion) as reported in \citet{Corsi2016}. The last is obtained by setting $t_e=55842$, $p=3$ \citep[as typically expected for Type Ib/c SNe; see][]{Chev06}, $\eta=10$, $\nu_{m,0}=1$\,GHz, and varying $r_0$, $\xi$, $\alpha_r$, $s$, and $B_0$. Model 0 is also plotted in  Figure~\ref{fig:lc_all_models_phase2} (solid line). 
Model 1 is a fit to the second radio peak ($t\gtrsim 587$ days since explosion) where we keep $r_0$ and $\alpha_r$ fixed to their best fit values for the first peak so as to ensure a smooth radial evolution between the first and second radio light curve peaks, and allow $\xi$, $s$, and $B_0$ to vary. This fit is similar to the one reported in  \citet{Palliyaguru2019} but updated to include the eMERLIN and HSA data presented here. 

\begin{figure}
    \begin{center}
    \includegraphics[width=9.cm,trim={ 2.5cm 12.5cm 1cm
2.0cm},clip]{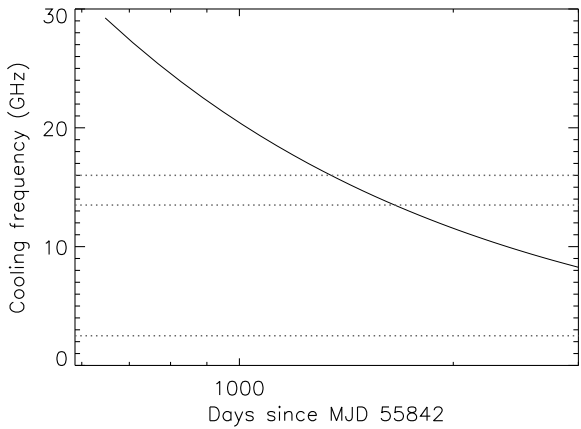}
\caption{Cooling frequency vs time during phase 2. The horizontal dotted lines mark the observation frequencies of 2.5 GHz, 13.5 GHz and 16 GHz.
For $\alpha_r=0.61$, $\nu_{m,0}=3 \, \rm GHz$ and $\eta=2$ (Model 3), $\nu_c$ (solid line) crosses $16 \, \rm GHz$ at $\sim 1345$ days since explosion. %
}
\label{fig:nuc}
\end{center}
\end{figure}

Compared to \citet{Palliyaguru2019}, the reduced $\chi^2$ for Model 1 is substantially higher, indicating a worsening of the goodness of fit. 
Model 2 is a fit with a model identical to Model 1 but where the 15\,GHz HSA data has been excluded. The improved $\chi^2$ value for Model 2 compared to Model 1 indicates a significant discrepancy between data and model at the highest radio frequencies, suggesting a steepening in the highest frequency light curve which may be caused by the passage of the cooling frequency in band. Indeed, the effects of synchrotron cooling may become important at the late timescales considered here. 
Within the SSA scenario, the synchrotron cooling frequency can be written as \citep[see Equation (A16) in ][]{skb05}:
\begin{equation} 
\nu_c= \frac{18\pi\,m_e\,c\,e}{(t-t_e)^2\sigma^2_T\,B^3}, 
\label{eq:coolfreq}
\end{equation}
where $\sigma_T$ is the Thomson cross-section and $e$ is the electron charge \citep{rl86,skb05}. 
For $\nu > \nu_c$, the flux density becomes \citep{skb05}: 
\begin{equation}
f_{\nu}(t) \propto \nu^{-p/2} \, t^{(6\alpha_r +(8-5p)\alpha_B+2(p-2)\alpha_{\gamma} -4p+2)/2}.
\end{equation}
Model 1 in Table~\ref{tb:model_results_ssa} predicts $\nu_c\approx50\,\rm GHz$ at $t-t_e\approx 7.5$\,yrs, which is above the highest frequency of our observations, resulting in large residuals with the HSA data point.
Thus, in Table~\ref{tb:model_results_ssa} we also report the results of a fit where we set $t_e=55842$, $p=3$, $s=0$, $\eta=2$, $\nu_{m,0}=3$\,GHz, $r_0=1.1\times10^{16}$\,cm, but allow $\alpha_r$, $B_0$, and $\xi$ to vary (Model 3).
As shown in Figure~\ref{fig:nuc}, with this choice the cooling frequency falls below 16\,GHz at $t-t_e\approx 1345$\,d ($\approx 3.6$\,yrs since explosion), thus causing a steeping of the light curve at this frequency. In Figure~\ref{fig:lc_all_models_phase2} we plot Model 3 with dotted lines, and the expected steeping ($f_{\nu}(t)\propto t^{-2.6}$ according to Equation (13)) of the 16\,GHz light curve due to synchrotron cooling with a dash-dot-dot-dotted line. Model 3 substantially improves the goodness of fit for the second radio peak compared to Models 1 and 2.

\begin{figure}
    \begin{center}
\includegraphics[width=9.cm,trim={ 2.5cm 12.5cm 1cm
2.0cm},clip]{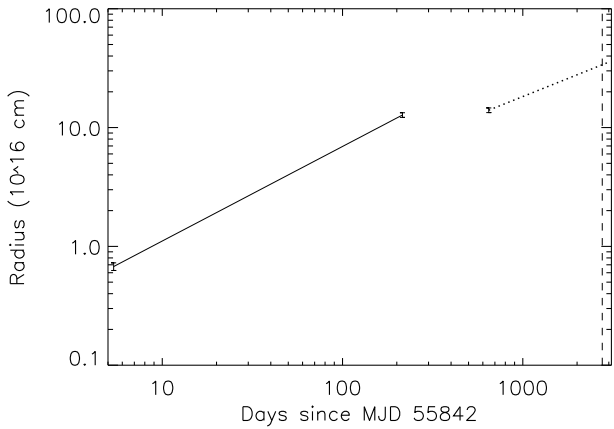}
\caption{
Radius vs time during phases 1 and 2. During phase 2, the radial evolution is modeled using Model 3 in table~\ref{tb:model_results_ssa}.
The date of the 16 GHz VLBI data is marked by the vertical dashed line. 
Our modeling does not include the time while encountering the CSM discontinuity and shows that the shock has decelerated during phase 2.
}
\label{fig:varyr0}
\end{center}
\end{figure}

In summary, while the SSA fits described in this Section have large limitations due to the simplifications that characterize the SSA model, overall they are indicative of the fact that (i) a non-constant CSM profile is needed to explain the PTF11qcj radio light curves, and (ii)  synchrotron cooling may be playing a role at late times. We also note that in Model 3 a flat ($s=0$) radial profile of the environment (ISM-like) is favored \citep[see][for a discussion of the $s=0$ case]{Chevalier1982}. Finally, as shown in 
Figure~\ref{fig:varyr0}, the combined radial evolution of the shock as implied by Model 0 for the first peak of the PTF11qcj light curves and by Model 3 for the second peak, implies that the shock has decelerated while encountering the CSM discontinuity.  
A similar case is of SN2014C, where VLBI observations revealed that the SN has substantially decelerated at late times after encountering a higher density shell \citep{Bietenholz2020}.

\subsection{SSA plus off-axis GRB scenario}
\label{sec:model_offGRB}
Similarly to what done in \citet{Palliyaguru2019}, we also use numerical simulations for off-axis GRB jets by \citet{ehm12} to model the second peak of PTF11qcj light curve within a constant density environment, considering the new HSA and eMERLIN data points. Within this scenario, the first radio peak is the radio emission from the SN ejecta while the second radio peak is due to the delayed emission from an initially off-axis jets that enters our line of sight after spreading and decelerating.
These two-dimensional hydrodynamic simulations of the GRB jets take into account the Blandford-Mckee solution \citep{bm76} in the relativistic regime, the Sedov-von Neumann--Taylor (SNT) solution \citep{t46}  in the late non-relativistic regime, and a transition in between regimes \citep{zm09}.

In Table~\ref{tb:model_results_offaxis} we report the best fit results 
for the isotropic equivalent kinetic energy of the explosion $E_{\rm iso}$, the jet half-opening angle, $\theta_0$, ISM density $n_{\rm ISM}$, and the observer angle, $\theta_{\rm obs}$,
when the latest four data points from HSA and eMERLIN, as well as the Chandra X-ray data point reported in \citet{Palliyaguru2019}, are added to the fit. 
These fits also assume equipartition of energy between particles and magnetic fields such that $\epsilon_e=\epsilon_B=0.33$. 
We set the electron energy index, $p=2.5$, which is typical for GRB afterglows, as expected from theoretical considerations \citep{Kirk2000,Achterberg2001}.
The best fit off-axis jet light curves are shown in Figure~\ref{fig:lc_all_models_phase2} (light blue, long-dashed lines).

\begin{table*}[htbp]
\centering
\begin{center}\begin{footnotesize}
\caption{Best fit results in the off-axis GRB scenario. The radius ($R_\perp$) and the corresponding angular diameter from modeling, and the expected radius from Equation~(\ref{eq:radius3}) ($R_{\perp,exp}$) and the corresponding angular diameter  are also listed. }
\label{tb:model_results_offaxis}
\begin{tabular}{lllllllllllllll}
\hline
\hline
Model&$E_{\rm iso}$ & $n$& $\theta_0$ &  $\theta_{\rm obs}$ & $p$ &$\chi^2$&$R_\perp$&angular size&$R_{\perp,exp}$ & expected angular size\\
\hline
&(erg)&($\rm cm^{-3}$)&(rad)&(rad)&&&(cm)&(mas)&(cm)&(mas)\\
\hline
Model 1&$1\times10^{53}$  & $1\times10^{-5}$   &  0.2& 0.4 &2.5 &$1285/41\approx31$&$3.0\times10^{19}$&35&$4.8\times10^{19}$&55\\
\hline
\end{tabular}
\end{footnotesize}\end{center}\end{table*}

\subsection{Size constraints: CSM-interacting vs off-axis GRB scenario}
\label{sec:size}
Within the SSA scenario, Models 0, 1, and 2 described in Section \ref{sec:model_SSA} all imply a shock radius around the time of the 16\,GHz HSA observation (2759 days post explosion) of $r\sim 10^{18}$\,cm. This in turn corresponds to an angular diameter of $\sim  1$ mas at the redshift $z=0.028$ of PTF11qcj, larger than the size constraints set by our HSA observations (see Table 1). On the other hand, Model 3 gives $r\approx 3\times10^{17}$ cm at 2759 days post explosion (see Figure 4), which corresponds to an angular diameter of $\approx 0.4$ mas, compatible with the HSA observations reported in Table 1. We note however that within the SSA model, which assumes spherical symmetry, we are not able to model any potential asymmetry in the shock geometry.

Within the off-axis GRB scenario for the second radio peak, the ejecta will no longer produce a symmetric image on the sky. Instead, the resolved image will highlight the front edge of the jet heading to the observer, producing an elongated and curved shape. 
For model 1 in Table~\ref{tb:model_results_offaxis}, this curved front will have traveled $R_{\perp,travel} = 3.0 \times 10^{19}$\,cm in projected distance from the origin of the explosion. The projected width of the image is $R_{\perp,w} = 1.4 \times 10^{19}$ cm, and its projected height $R_{\perp,h} = 2.5 \times 10^{18}$ cm.
These correspond to a projected angular distance from the origin of the explosion of 35\,mas, a projected angular width of 8\,mas, and a projected angular height of 14\,mas.  Thus, if the second radio light curve peak of PTF11qcj was due to an off-axis GRB, we should have resolved a much larger image in our HSA observations.

We note that while the off--axis GRB model considered here assumes a top-hat jet, a structured jet such as the one considered in \citet{np17} would imply sizes that can be bracketed by the the two extreme cases of a spherically symmetric blast wave, and a non-spreading relativistic cone.
 Indeed, for a spherically symmetric Sedov-Taylor blast wave, the analytical expression for the shock radius at a late--time $t$ is given  by \citep{vaneerten2018}:
\begin{eqnarray}
    \nonumber R=C_R(k)\left(\frac{E_j}{10^{51} \rm erg}\right)^{\frac{1}{5-k}}\left(\frac{\rho_{ref}}{m_p}\right)^{\frac{-1}{5-k}}\times\\\left(\frac{R_{ref}}{c\times10^{6}}\right)^{\frac{-k}{5-k}}\left(\frac{t}{10^{8}}\right)^{\frac{2}{5-k}}
\end{eqnarray}
where $C_R(k)=1.5$ parsec for the power law index of the CSM density profile $k=1.4$ (Model 1, Table~\ref{tb:model_results_ssa}), 
$E_j$ is the total energy in the ejecta, $\rho_{\rm ref}$ and $R_{\rm ref}$ are the circumburst medium density and radius at a reference time (which we set to 10 days), respectively \citep{vaneerten2018}. For the best fit parameters of Model 1 in Table~\ref{tb:model_results_ssa}, we find the expected radius would be $8.9\times10^{17} \rm cm\, (1 \,\rm mas)$, larger than our HSA constraint. 

For a relativistic jet expanding into the ISM, the apparent radius at time $t$ may be calculated using the analytical expression \citep{Oren2004}:
\begin{equation}
    R_{\perp,exp}=5\times10^{16}\left(\frac{E_{51}}{n}\right)^{1/6}T_j^{-1/8}t^{5/8},
    \label{eq:radius3}
\end{equation}
where $T_j=(E_{51}/n)^{1/3}(\theta_0/0.1)^2(1+z)$ is the jet break time, $\theta_0$ is the jet half opening angle, $n$ is the ISM number density \citep{Oren2004}.
For the best fit parameters of $E_{\rm iso}$, $\theta_0$, $n_{\rm ISM}$ listed in Table~\ref{tb:model_results_offaxis}, the expected radius at $2759$ days post explosion 
is $4.8\times10^{19}$ cm.
This corresponds to an angular size of $\sim 55$ mas at the redshift $z=0.028$ of PTF11qcj, much larger than our HSA constraint. 

Based on these results, the off-axis hypothesis for the origin of the second radio peak of the PTF11qcj light curves is disfavored.

\section{Summary and Conclusion}
\label{sec:conc}
We have presented HSA and eMERLIN observations of PTF11qcj obtained $\sim$7.5 yrs post explosion.
The source is marginally resolved in the HSA data at 15 GHz, indicating a diameter of $(300 \pm 71) \,\rm \mu as$.
This corresponds to an average expansion velocity of $\approx0.036\pm0.008\,c$.

Modeling of the light curves within an SSA scenario requires the interaction of the shock with a non-smooth CSM whose radial profile changes from a stellar wind profile ($n_e\propto r^{-2}$) to a constant density medium. This variable CSM profile also affects the temporal evolution of the shock radius.  
We also find that synchrotron cooling may be playing a role at the highest radio frequencies.
Within this SSA model with variable CSM, the derived size of the shock at the time of our HSA observations can be reconciled with our measurements. Our results are thus consistent with the CSM interaction model (considering that such model is highly simplified). The size constraints derived from our HSA observations also seem to disfavor an off-axis relativistic jet scenario for the radio re-brightening of PTF11qcj.

In conclusion, the VLBI observations reported in this paper favor the original interpretation of PTF11qcj as a strongly CSM-interacting radio SN \citep[see][]{Corsi2014,Palliyaguru2019}. However, our conclusions are limited by the simplifications inherent in the spherically symmetric SSA model adopted here. We encourage more detailed theoretical modeling aimed at interpreting PTF11qcj complex radio light curve and available VLBI data. 

\acknowledgments
A.C. acknowledges support from the National Science Foundation CAREER Award \#1455090, and from the Chandra GO Award \#GO7-18065X. 
N.T.P acknowledges support from NSF NANOGrav Physics Frontier Center (NSF Grant No. PFC-1430284) and start-up funds to J.D. Romano from TTU.
M.P.-T. acknowledges financial support from the State Agency for Research of the Spanish Ministry of Science, University, and Education (MCIU) through the "Center of Excellence Severo Ochoa" award for the Instituto de Astrof\'isica de Andaluc\'ia (SEV-2017-0709) and through grants AYA2015-63939-C2-1-P and  PGC2018-098915-B-C21.
The National Radio Astronomy Observatory is a facility of the National Science Foundation operated under cooperative agreement by Associated Universities, Inc.
e-MERLIN is a National Facility operated by the University of Manchester at Jodrell Bank Observatory on behalf of STFC.
This work made use of the Swinburne University of Technology software 
correlator, developed as part of the Australian Major National Research 
Facilities Programme and operated under licence. 

\bibliographystyle{apj}

\bibliography{bibliography1}

\end{document}